\documentclass{article}
\usepackage{spconf,amsmath,graphicx}

\usepackage{microtype}
\usepackage{graphicx}
\usepackage{subfigure}
\usepackage{booktabs} 

\usepackage{enumerate}
\usepackage{subfigure}
\usepackage{cite}
\usepackage{multirow}

\usepackage{hyperref}
\usepackage[dvipsnames]{xcolor}

\usepackage{amsmath,amsfonts,bm}

\def\eqref#1{equation~\ref{#1}}

\def\1{\bm{1}}

\def\rw{{\textnormal{w}}}

\def\vh{{\bm{h}}}

\def\vl{{\bm{l}}}

\def\vp{{\bm{p}}}
\def\vq{{\bm{q}}}

\def\vs{{\bm{s}}}

\def\vv{{\bm{v}}}
\def\vw{{\bm{w}}}
\def\vx{{\bm{x}}}
\def\vy{{\bm{y}}}
\def\vz{{\bm{z}}}

\def\evl{{l}}

\def\evp{{p}}
\def\evq{{q}}

\def\evs{{s}}

\def\evv{{v}}

\def\evx{{x}}

\def\mH{{\bm{H}}}

\DeclareMathAlphabet{\mathsfit}{\encodingdefault}{\sfdefault}{m}{sl}
\SetMathAlphabet{\mathsfit}{bold}{\encodingdefault}{\sfdefault}{bx}{n}

\newcommand{\E}{\mathbb{E}}

\definecolor{amethyst}{rgb}{0.6, 0.4, 0.8}

\newcommand{\model}{AutoTTS}

\title{\textsc{AutoTTS}\uppercase{: End-to-End Text-to-Speech Synthesis through\\ Differentiable Duration Modeling}}
\name{Bac Nguyen, Fabien Cardinaux, Stefan Uhlich}
\address{Sony Europe B.V., R\&D Center, Stuttgart Laboratory 1, Germany}
\begin{document}

\ninept
\setlength{\abovedisplayshortskip}{1ex plus1ex minus1ex}
\setlength{\abovedisplayskip}{1ex plus1ex minus1ex}
\setlength{\belowdisplayshortskip}{2ex plus1ex minus1ex}
\setlength{\belowdisplayskip}{2ex plus1ex minus1ex}
\maketitle
\begin{abstract}
 Parallel text-to-speech (TTS) models have recently enabled fast and highly-natural speech synthesis. However, they typically require external alignment models, which are not necessarily optimized for the decoder as they are not jointly trained. In this paper, we propose a differentiable duration method for learning monotonic alignments between input and output sequences. Our method is based on a soft-duration mechanism that optimizes a stochastic process in expectation. Using this differentiable duration method, we introduce AutoTTS, a direct text-to-waveform speech synthesis model. AutoTTS enables high-fidelity speech synthesis through a combination of adversarial training and matching the total ground-truth duration. Experimental results show that our model obtains competitive results while enjoying a much simpler training pipeline. Audio samples are  available online\footnote{Audio samples are available at \url{https://sony.github.io/ai-research-code/autotts/demo/}}.
\end{abstract}

\section{Introduction}

Text-to-speech (TTS) or speech synthesis is the process of converting written text into speech. Recently, rapid progress has been made in this area, which enables TTS models to synthesize natural and intelligible speech. Many TTS systems are based on autoregressive models such as Tacotron~\cite{wang2017tacotron,8461368}, Deep Voice~\cite{pmlr-v70-arik17a,10.5555/3294996.3295056}, and Transformer TTS~\cite{li2019neural}. The idea is to learn an encoder-decoder model using a soft attention mechanism. Given a sequence of characters or phonemes as inputs, the encoder learns a sequence of hidden representations, which are passed to the decoder to predict the acoustic parameters, such as fundamental frequencies and (mel-)spectrograms. In addition, a neural vocoder such as WaveNet~\cite{vandenOord_2016}, Parallel WaveGAN~\cite{9053795}, or HifiGAN~\cite{NEURIPS2020_c5d73680} is used to synthesize the speech waveform from the output of the acoustic model. Despite their satisfactory performance, autoregressive acoustic TTS models typically suffer from robustness (\textit{e.g.,} word repetitions and word skipping) and slow inference speed~\cite{NEURIPS2019_f63f65b5}. The latter limits their applicability in real-time systems that require fast speech generation.

\begin{figure}[ht!]
    \centering
    \includegraphics[width=0.8\linewidth]{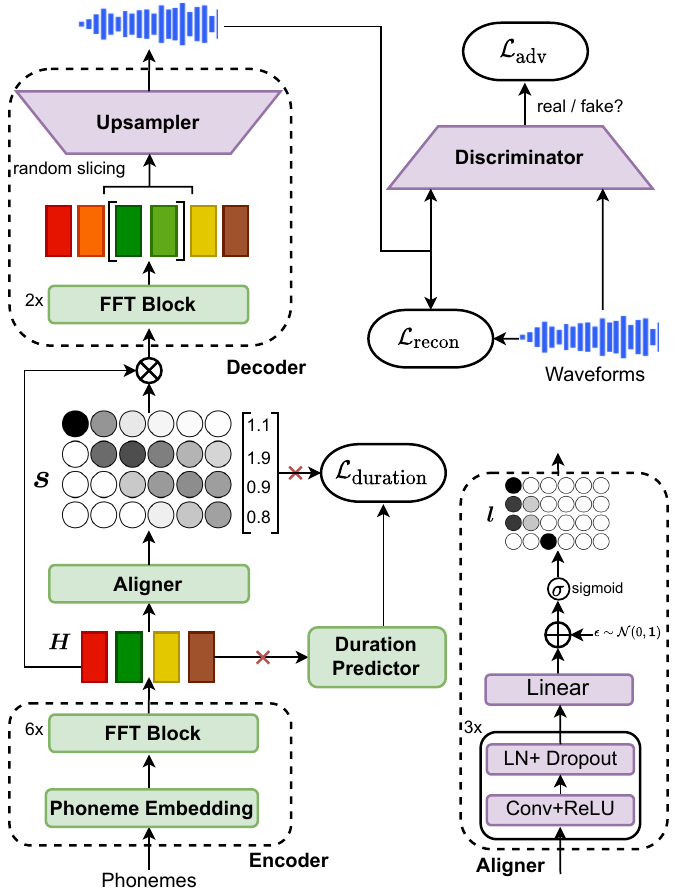}
    \caption{The overall architecture and training procedure of \model{}. During training, audio waveforms of fixed length are generated by randomly selecting a sliced hidden sequence. The red cross mark indicates the stop gradient operation.} 
    \label{fig:overview}
    \vspace{-0.5cm}
\end{figure}

To speed up the inference, parallel (or non-autoregressive) TTS models have been proposed~\cite{pmlr-v119-peng20a,ren2020fastspeech,NEURIPS2019_f63f65b5,9414718,elias21_interspeech,popov2021grad}. These models can fully leverage the parallel computation while obtaining reasonably good results. However, learning alignments between text and its corresponding mel-spectrogram becomes challenging for non-autoregressive models. The alignments must satisfy several important criteria~\cite{he2019robust}: (1) each character/phoneme token should be aligned with at least one spectrogram frame, (2) the aligned spectrogram frames should be consecutive, and (3) the alignment should be monotonic. One common solution is to rely on the attention outputs of a pretrained autoregressive teacher model to extract the alignments~\cite{NEURIPS2019_f63f65b5}. Alternatively, DurIAN~\cite{Yu2020} and FastSpeech~2~\cite{ren2020fastspeech} used token durations extracted by forced alignment~\cite{mcauliffe2017montreal} to obtain alignments. The sequence of hidden representations is expanded according to the duration of each token to match the length of the mel-spectrogram sequence. However, these models are naturally sensitive to the performance of the teacher or the external aligner. Therefore, it is important to design a TTS model which can directly learn the alignments from data, which we propose in this paper. 

Despite the significant progress achieved in parallel TTS, the distribution mismatch between the acoustic model and the vocoder often leads to artifacts in synthesized speech~\cite{donahue2021endtoend,kim2021conditional}. To obtain high-quality speech, two-stage TTS models often require sequential training or fine-tuning. That is, the vocoder is trained with the output generated by the acoustic model. Due to the dependency on the vocoder, the performance of two-stage models is limited. This has motivated the development of fully end-to-end TTS models. The most obvious advantages of end-to-end models include (1) avoiding the error propagation from intermediate steps and (2) reducing the complexity of the training and deployment pipeline. Nevertheless, training an end-to-end TTS system is a difficult task due to the huge difference between text and audio waveform modalities. 

In this paper, we introduce \model{} (see Fig.~\ref{fig:overview}), a parallel end-to-end TTS model, which generates audio waveforms directly from a sequence of phonemes. The contributions of our work are summarized as follows. (i) We propose a novel method for duration modeling. Our differentiable duration model allows back-propagation of gradients through the entire network. As a result, it enables \model{} to learn the alignments between phoneme sequences and audio waveforms without additional external aligners. (ii) Using this differential duration model, we further improve the perceptual audio quality of AutoTTS by leveraging the adversarial training in an end-to-end fashion and by matching the total ground-truth duration. (iii) We  experimentally validate the effectiveness of AutoTTS in terms of  speech  quality and  inference speed against other multi-stage training TTS models.



\section{Related work}
%
\textbf{Alignment models.} Recent TTS models have exploited the duration prediction of each character/phoneme token to obtain alignments between text and speech.
During training, the token durations can be extracted from autoregressive TTS models~\cite{NEURIPS2019_f63f65b5,VainerD20} or external aligner models~\cite{ren2020fastspeech,Yu2020,9414718}. Another approach is to explicitly train an internal aligner to extract durations. For instance, AlignTTS~\cite{DBLP:conf/icassp/ZengWCXX20} considered all possible alignments to align text to the mel-spectrogram by using dynamic programming with multi-stage training. JDI-T~\cite{lim2020jdi} avoided multi-stage training by jointly training both autoregressive and non-autoregressive models to get the token durations. EATS~\cite{donahue2021endtoend} and Parallel Tacotron~2~\cite{elias21_interspeech} used the soft dynamic time warping as a reconstruction loss to optimize the duration prediction. Note that these methods directly learn the duration of each token, whereas our method models duration as an expectation over binary random variables. EfficientTTS~\cite{pmlr-v139-miao21a} introduced monotonic constraints to the alignments using an index mapping vector. GlowTTS~\cite{NEURIPS2020_5c3b99e8} proposed hard monotonic alignments, which can be obtained efficiently via dynamic programming. Compared to our duration proposal, previous methods are either computationally expensive or hard to implement.

\textbf{Fully end-to-end TTS.} To further improve intelligibility and naturalness of synthesized speech, some efforts have been made to develop a fully end-to-end TTS system for direct text to waveform synthesis. For instance, FastSpeech 2s~\cite{ren2020fastspeech} incorporated a mel-spectrogram decoder to improve the linguistic embeddings and adopted adversarial training to generate audio waveforms.  EATS~\cite{donahue2021endtoend} employed differentiable alignments and adversarial training in an end-to-end manner. VITS~\cite{kim2021conditional} adopted variational inference with normalizing flow and adversarial training. Compared to these approaches, our model \model{} enjoys a much simpler training pipeline, while still being end-to-end.

\section{Proposed method}

In this section, we present a novel method called \model{} that enables prediction of the raw audio waveform from a sequence of phonemes. The overall architecture and training procedure are illustrated in Fig.~\ref{fig:overview}. Essentially, \model{} consists of an encoder, an aligner, and a decoder network. The encoder maps a sequence of phonemes into some hidden embeddings. The aligner maps these hidden embeddings to other embeddings which are aligned with the audio output but in a lower resolution. The decoder then upsamples the output embeddings from the aligner to the raw audio waveform. Our method does not require supervised duration signals, which enables back-propagation-based training. In the following, we first introduce our duration modeling to align the text and speech. Then, we describe the adversarial training, followed by some network architecture and implementation details of \model{}.

\subsection{Duration modeling} \label{sec:stochastic}
We begin by reviewing the length regulator used in FastSpeech~\cite{NEURIPS2019_f63f65b5} for solving the length mismatch problem between the phoneme sequence and mel-spectrogram output. Given a sequence of $N$ phonemes $\vx=\{\evx_1, \dots, \evx_N\}$, the encoder outputs  a sequence of hidden states $\mH=\{\vh_1, \dots, \vh_N\}$ where $\vh_i \in \mathbb{R}^D$ for $i=1,\dots, N$. The decoder then outputs a sequence of mel-spectrogram frames for a given sequence of hidden states. However, the length of the phoneme sequence is often much smaller than that of the mel-spectrogram. To resolve this problem, a length regulator is used to upsample each hidden state according to its  duration. Here, the duration is defined as the number of mel-spectrogram frames corresponding to that phoneme. During training, durations are extracted by an external aligner~\cite{mcauliffe2017montreal} or teacher-student distillation~\cite{NEURIPS2019_f63f65b5}. Since the length regulator is a non-differentiable function, gradient-based methods cannot be used to find the optimal durations. We will address this issue with a stochastic duration model, which enables end-to-end training.

We formulate the duration as a stochastic process. Assume that the duration of a phoneme is a discrete integer in the range of $[0, M]$. In particular, let $\rw_i$ be a random variable indicating the duration of the $i$-th phoneme and $\vp_i \in [0, 1]^{M}$ be a vector containing the parameters of the distribution that characterizes $\rw_i$, \textit{i.e.,} $\rw_i \sim P(\rw_i| \vp_{i})$. The probability of having a duration of $m \in \{1, \dots, M\}$ for the $i$-th phoneme is defined as
\begin{align}
    \evl_{i, m} \!=\! \evp_{i,m} \!\!\! \prod_{k=1}^{m - 1}\!\! (1 - \evp_{i,k}) \!=\! \evp_{i, m} \texttt{cumprod}(1 - \vp_{i,:})_{m - 1}, \label{eq:l}
\end{align}
where $\texttt{cumprod}(\vv) = [\evv_1, \evv_1\evv_2, \dots,  \prod_{i=1}^{|\vv|}\evv_i]$ is the cumulative product operation. We refer to $\vl$ as the length probability. Given the parameters $\vp_i$, one can sample the duration following a sequence of $\text{Bernoulli}(\evp_{i,m})$ distributions, starting from $m=1$. As soon as we obtain an outcome being one for some $m$, we stop and set the duration to $m$. If there is no outcome after $M$ trials, the duration is set to zero\footnote{This is used to model special characters such as punctuation marks of the input text. It does not contradict the assumptions of alignments.} with a probability of
\begin{align*}
    \evl_{i, 0}= \prod_{k=1}^{M} (1 - \evp_{i,k}) =  \texttt{cumprod}(1 - \vp_{i,:})_M\,.
\end{align*}
It is easy to show that $\sum_{m=0}^M \evl_{i, m} = 1$, for every $i \in \{1, \dots, N\}$. In other words, the length probability is a valid probability distribution.

The duration of several phonemes can be summed over the duration of each individual phoneme. Let $\evq_{i,j}$ denote the probability that a sequence of the first $i$ phonemes having a duration of $j\in \{0, \dots, M\}$. For the first phoneme, this is identical to the length probability, \textit{i.e.}, $\vq_{1,:} = \vl_{1, :}$. For $i > 1$, $\vq_{i, :}$ can be recursively formulated with $\vq_{i-1,:}$  and $\vl_{i, :}$ as in Eq.~(\ref{eq:q}) by considering all possible durations of the $i$-th phoneme, \textit{i.e.},
\begin{align}
    \evq_{i,j} = \sum_{m=0}^{j} \evq_{i - 1,m} \evl_{i,j - m} \,. \label{eq:q}
\end{align}

\begin{figure}[t]
     \centering
     \hfill
     \subfigure[Length probability $\vl$]{\includegraphics[width=0.48\linewidth]{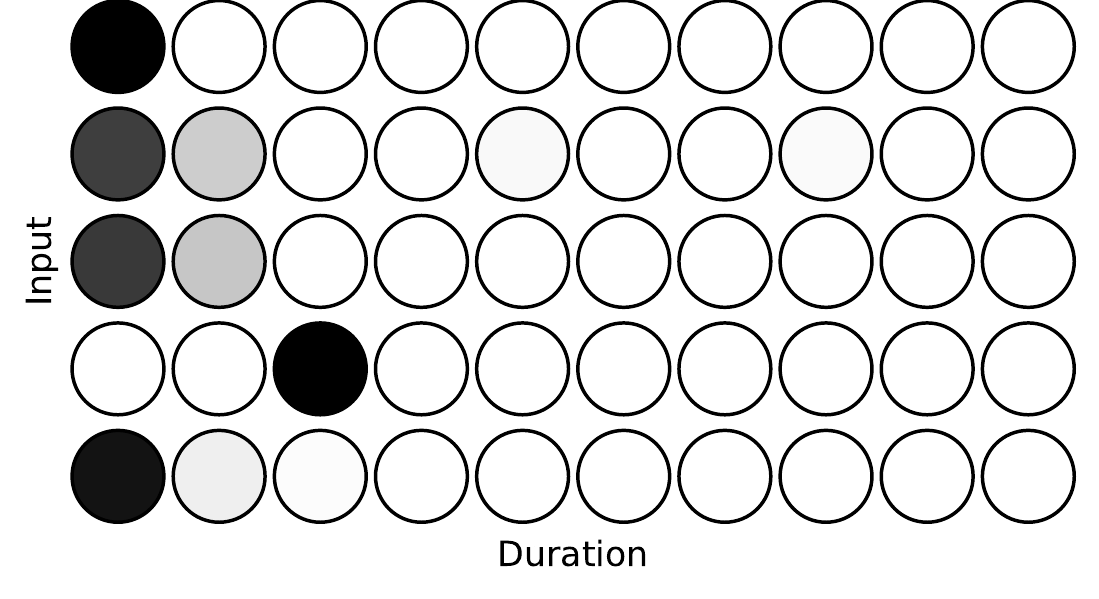}}
     \hfill
     \subfigure[Attention probability $\vs$]{\includegraphics[width=0.48\linewidth]{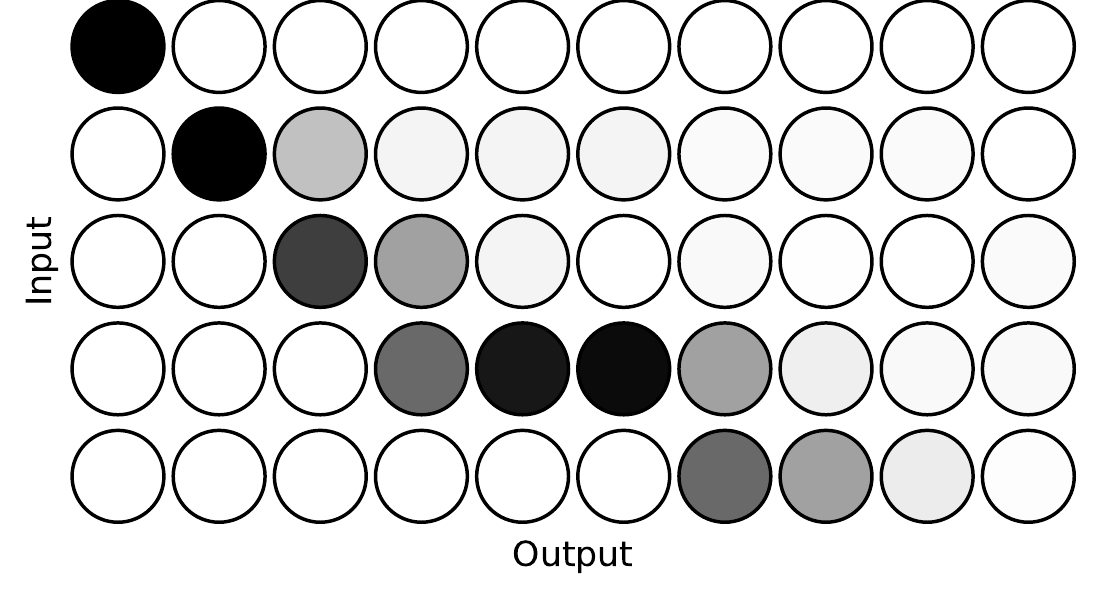}}
     \hfill
    \caption{Illustration of our duration model using our stochastic process from Subsection~\ref{sec:stochastic}.}
    \label{fig:stochastic}
    \vspace{-0.5cm}
\end{figure}

After having the duration probability of a sequence of phonemes, we need to compute the attention or alignment probability. Let $\evs_{i,j}$ denote the probability that the $j$-th output frame is aligned with the $i$-th input token. The alignment between $\vh_{i,:}$ and $\vy_{j,:}$ occurs when the total duration of the first $i$ phonemes is bigger than or equal to $j$. On the first phoneme, we can compute $ \evs_{1, j} = \sum_{m=j}^{M} \evl_{i,m}$. For $i > 1$, $\vs_{i,:}$ can be computed with $\vq_{i-1,:}$ and $\vl_{i,:}$ as
\begin{align}
    \evs_{i,j} \!=\!\!\! \sum_{m=0}^{j - 1}\!\!\evq_{i-1, m}\!\!\! \!\!\sum_{k=j - m}^{M} \!\!\!\!\!\evl_{i,k} \!\!= \!\!\sum_{m=0}^{j - 1}\!\evq_{i-1, m}  \texttt{cumsum}^{*}(\vl_{i,:})_{j-m} , \label{eq:s}
\end{align}
where $\texttt{cumsum}^{*}(\vv) = [\sum_{i=1}^{|\vv|}\evv_i,\sum_{i=2}^{|\vv|}\evv_i, \dots, \evv_{|\vv|}]$ is the reverse cumulative sum operation. Here we consider all possibilities that the $i$-th phoneme aligns the $j$-th output frame. In Fig.~\ref{fig:stochastic}, we illustrate how the attention probability $\vs$ is derived from the length probability $\vl$. By modeling the duration using cumulative summation, we implicitly enforce the alignment being monotonic. After computing the attention probability $\vs$, we can upsample the hidden states as expected output $\E [\vy_{j,:}] = \sum_{i=1}^{N} \evs_{i,j}\vh_{i,:}$, for $j=\{1,\dots,M\}$.

\subsection{Training procedure}
\model{} is trained based on an adversarial learning scheme~\cite{NIPS2014_5ca3e9b1}. A discriminator $D$ is used to distinguish between the real speech waveform $\vz$ and the audio waveform produced by our network $G(\vx)$. In particular, the following loss function is used to train \model{}
\begin{align*}
 \mathcal{L} = \mathcal{L}_{\text{adv-G}} + \lambda_{\text{length}}  \mathcal{L}_{\text{length}} + \lambda_{\text{duration}} \mathcal{L}_{\text{duration}} +  \lambda_{\text{recon}} \mathcal{L}_{\text{recon}} \,,
\end{align*}
where $\mathcal{L}_{\text{adv-G}}$, $\mathcal{L}_{\text{length}}$, $\mathcal{L}_{\text{duration}}$, and $ \mathcal{L}_{\text{recon}}$ indicate the adversarial, length, duration, and reconstruction losses, respectively; $\lambda_{\text{length}}, \lambda_{\text{duration}}$, and $\lambda_{\text{recon}}$ are weighing terms. The discriminator is simultaneously trained using the adversarial loss $\mathcal{L}_{\text{adv-D}}$.The detail of each loss function will be described in the following. 

\textbf{Adversarial loss.} The least-squares loss~\cite{Mao_2017_ICCV} is employed for adversarial training, \textit{i.e.,}
\begin{align*}
    \mathcal{L}_{\text{adv-D}} &= \E_{(\vx, \vz)} \left[ (D(\vz) - 1)^2 + D(G(\vx))^2\right] \,, \\ 
    \mathcal{L}_{\text{adv-G}} &= \E_{\vx} \left[ (D(G(\vx)) - 1)^2\right]\,.   
\end{align*}
On one hand, the discriminator forces the output of real samples to be one and that of synthesized samples to be zero. On the other hand, the generator is trained to fool the discriminator by producing samples that will be classified as real samples. This training scheme helps to synthesize realistic speech.

\textbf{Length loss.} Based on the length probability, the expected duration of the $i$-th phoneme can be computed as
\begin{align*}
    \E_{\rw_i \sim P(\rw_i\mid \vp_{i})}[\rw_i] = \sum_{m = 1}^{M} m \evl_{i, m}\,.
\end{align*}
The expected length of the entire utterance is computed by summing up all phoneme duration predictions. We encourage this expected length to be close to the ground truth length $M_{\text{total}}$ of the speech by minimizing the following loss
\begin{align*}
    \mathcal{L}_{\text{length}} = \frac{1}{N}\left| M_{\text{total}} - \sum_{i=1}^N \E_{\rw_i \sim P(\rw_i\mid \vp_{i})}[\rw_i] \right| \,.
\end{align*}

\textbf{Duration loss.} To speed up the inference, a duration predictor $f$ is used to estimate the phoneme durations. More concretely, the duration predictor takes phoneme hidden sequences $\vh_i$ as inputs and takes durations extracted from our aligner as targets. During training, we stop the gradient propagation from the duration predictor to the encoder and the aligner. Our duration loss is summarized as
\begin{align*}
    \mathcal{L}_{\text{duration}} = \frac{1}{N}\sum_{i=1}^N\left| f(\text{sg}[\vh_i]) -  \text{sg}\Big[\E_{\rw_i \sim P(\rw_i\mid \vp_{i})}[\rw_i] \Big] \right| \,,
\end{align*}
where $\text{sg}[.]$ indicates the stop gradient operator. 
Note that the outputs of the duration predictor are discretized to the closest frame. In order to allow the duration predictor and the aligner to converge to similar outputs, the aligner is trained to encourage discrete outputs (see Subsection \ref{sec:arch}). Using this approach,we empirically observe that there is no performance drop when using the duration predictor instead of the aligner during inference.

\textbf{Reconstruction loss.} Given a sequence of phonemes, our network should be able to reconstruct the corresponding speech. To this end, the feature matching loss~\cite{10.5555/3045390.3045555} and the spectral loss~\cite{NEURIPS2020_c5d73680} are adopted. In particular, we force the synthesized speech to be as similar as the real speech by minimizing
\begin{align*}
    \mathcal{L}_{\text{recon}} &= \E_{(\vx, \vz)} \left[ \sum_{t=1}^T \| D^{(t)}(G(\vx)) - D^{(t)}(\vz) \|_1 \right] \\
    & + \lambda_{\text{mel}}\E_{(\vx, \vz)} \left[ \| \phi(G(\vx)) - \phi(\vz) \|_1 \right] \,,
\end{align*}
where $D^{(t)}$ is the feature map output from the discriminator $D$ at the $t$-th layer,  $\phi$ is the log-magnitude of mel-spectrogram, and $\lambda_{\text{mel}}$ is a weighing term.

\subsection{Network architecture and efficient implementation} \label{sec:arch}
In this section, we describe the neural network architecture of \model{}. The encoder network is a transformer-based encoder~\cite{NIPS2017_3f5ee243}. It consists of a stack of six Feed-Forward Transformer (FFT) blocks as in FastSpeech~\cite{NEURIPS2019_f63f65b5}. Each FFT block includes self-attention and 1D-convolutional layers of kernel size 9. For the self-attention mechanism, we use the relative positional representation~\cite{shaw-etal-2018-self}. The decoder network consists of two FFT blocks and the Upsampler network.  The decoder aims to upsample the output sequence of the aligner to match the temporal resolution of the raw audio waveform. Upsampler is a fully-convolutional neural network, which is inherited from HiFi-GAN~\cite{NEURIPS2020_c5d73680}. We empirically observe that without the FFT blocks in the decoder, training converges very slowly. These FFT blocks help to capture the longer-term dependencies of the data. The aligner network consists of three 1D-convolutional layers of kernel size 5 with the ReLU activation, followed by layer normalization and dropout (see Fig.~\ref{fig:overview}). A linear layer is added to project the hidden states into a vector of size $M$ containing the parameters of the distribution that characterizes~$\vw_i$. The duration predictor shares a similar architecture like that of the aligner, except that the last linear layer outputs a single scalar, indicating the phoneme duration. We adopt the discriminator architecture as in~\cite{NEURIPS2020_c5d73680}, which consists of several multi-period discriminators and multi-scale discriminators operating on different resolutions of the input.

To make our method more efficient and stable to train, we use the following implementation for \model{}. The cumulative product in Eq.~(\ref{eq:l}) is numerically unstable for the gradient computation. We resolve this issue by computing this product in the log-space. The computation of the probability matrix $\vq$ in Eq.~(\ref{eq:q}) and $\vs$ in Eq.~(\ref{eq:s}) can be computationally expensive. Fortunately, we can efficiently implement them as convolution operations, which enjoy computational benefits from parallel computing. Another issue is that the attention probability matrix $\vs$ might not produce hard alignments as our length probability also produces soft outputs. Ideally, we would like to have discrete durations, which enable the alignments with the length regulator at inference time. To encourage the discreteness, we simply add zero-mean and unit-variance Gaussian noise before the sigmoid function which produces $\vl$ as in~\cite{SALAKHUTDINOV2009969}.

\section{Experiments}
\subsection{Experimental setups}
We evaluate our model \model{} on the LJSpeech data set~\cite{ljspeech17}, which consists of 13,100 English audio clips at the sampling rate of 22,050. LJSpeech contains 24 hours of high-quality speech data with a single speaker.  The data set is split into two sets, a set of 12,588 samples for training and another set of 512 samples for testing. Texts are normalized to sequences of phonemes using phonemizer~\cite{Bernard2021}. To better model the prosody, all punctuation marks are preserved in the output of text normalization.

Our model is trained for 3,000 epochs with a batch size of 22 using 8 NVIDIA A100 GPUs. Training takes about 1.8 minutes per epoch. We use the AdamW otpimizer~\cite{loshchilov2017decoupled} with $\beta_1=0.8$, $\beta_2=0.99$, and a learning rate of $2\times 10^{-4}$. During training, we randomly extract 128 frames ($\sim1.5$ seconds) of the hidden representation to feed to Upsampler to alleviate the memory constraint on the GPUs. Training targets are then defined as the corresponding audio segments extracted from the ground truth audio waveforms.

We compare the mean opinion score (MOS) of audio generated by \model{} with other systems, including Tacotron~2~\cite{8461368}, FastSpeech~2~\cite{ren2020fastspeech}, HiFi-GAN+Mel (where the ground truth audio are converted to mel-spectrograms, then converted back to audio waveforms using the pre-trained HiFi-GAN vocoder~\cite{NEURIPS2020_c5d73680}), and the ground-truth audio. For a fair comparison, we also use HiFi-GAN as the vocoder  for Tacotron 2 and  FastSpeech 2. All audio samples for the MOS study are generated by randomly choosing transcripts from the test set. We normalize all audio waveforms  to avoid the effect of amplitude differences. The subjective test was conducted internally by 20 participants. For each utterance, each listener assigns a score in a five-point Likert scale (where 1: Bad, 2: Poor, 3: Fair, 4: Good, and 5: Excellent) with a rating increment of 0.5. 

\subsection{Experimental results}
The subjective evaluation results shown in Table~\ref{tab:mos} shows that \model{} achieves significantly better opinion scores than the two-stage models, namely Tacotron~2 and FastSpeech~2. This is remarkable since other end-to-end TTS systems such as FastSpeech 2s~\cite{ren2020fastspeech} and EATS~\cite{donahue2021endtoend} perform worse than two-stage TTS as reported in their studies. Note that the recently introduced NaturalSpeech end-to-end TTS~\cite{tan2022naturalspeech} also outperforms two-stage models but at the cost of a more complex training pipeline (phoneme pre-training). Note also that \model{}'s score is only slightly below the ground-truth audio score.


\begin{table}[h]

  \caption{MOS with 95\% confidence intervals (CI)}
  \label{tab:mos}
  \centering
  \begin{tabular}{lrr}
    \toprule
    \textbf{Method} & \textbf{MOS} & \textbf{CI}\\
    \midrule
    Ground Truth & 4.81& $\pm$0.04\\
    HiFi-GAN+Mel & 4.76& $\pm$0.05\\
    \midrule
    Tacotron 2 & 3.92 & $\pm$0.07\\
    FastSpeech 2& 3.75& $\pm$0.07\\
    \midrule
    \model{} & \textbf{4.28}& \textbf{$\pm$0.06}\\
    \bottomrule
  \end{tabular}
  \vspace{-0.5cm}
\end{table}

Furthermore, we report the number of model parameters (in million) and inference speed of the competing methods in Table~\ref{tab:model}. The last two columns denote the real time (in seconds) required to synthesize ten seconds of speech waveforms on CPU and GPU, respectively. Real time is measured on an Intel(R) Xeon(R) CPU @ 3.40GHz with one NVIDIA GTX 1080 Ti GPU.  Additionally, the vocoder parameters are listed for Tacotron 2 and FastSpeech 2 because both need a vocoder for speech generation. \model{} has slightly more parameters compared to these acoustic models since they output mel-spectrograms instead of raw audio waveforms. We can see that \model{} is faster than other competing methods on both CPU and GPU due to the fully end-to-end generation.

\begin{table}[h]
\vspace{-0.5cm}
  \caption{Comparison of model size and inference speed}
  \label{tab:model}
  \centering
  \setlength{\tabcolsep}{5pt}
  \begin{tabular}{p{1.7cm}rrcrr}
    \toprule
    \multirow{2}{*}{\textbf{Method}} & \multicolumn{2}{c}{\textbf{\# of parameters} (M)} & & \multicolumn{2}{c}{\textbf{Real time} (s)} \\
    \cmidrule{2-3} \cmidrule{5-6}
    & \textbf{Model} & \textbf{Vocoder} & &\textbf{CPU} & \textbf{GPU}\\
    \midrule
    Tacotron 2 & 28.19& 13.92 & & 8.48 & 1.72\\
    FastSpeech 2& 35.16 & 13.92 & & 5.14 & 0.21\\
    \midrule
    \model{} & 38.69 & 0.00& & \textbf{1.77} &\textbf{ 0.12} \\
    \bottomrule
  \end{tabular}
\end{table}

\begin{figure}[tb]
     \centering
     \hfill
     \subfigure[Attention matrix]{\label{fig:alignment:att}\includegraphics[width=0.48\linewidth]{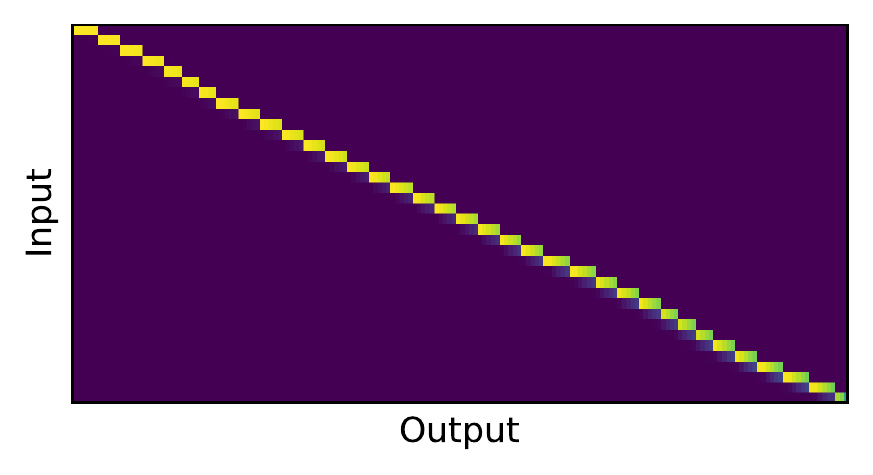}}
     \hfill
     \subfigure[Alignment]{\label{fig:alignment:mel}\includegraphics[width=0.48\linewidth]{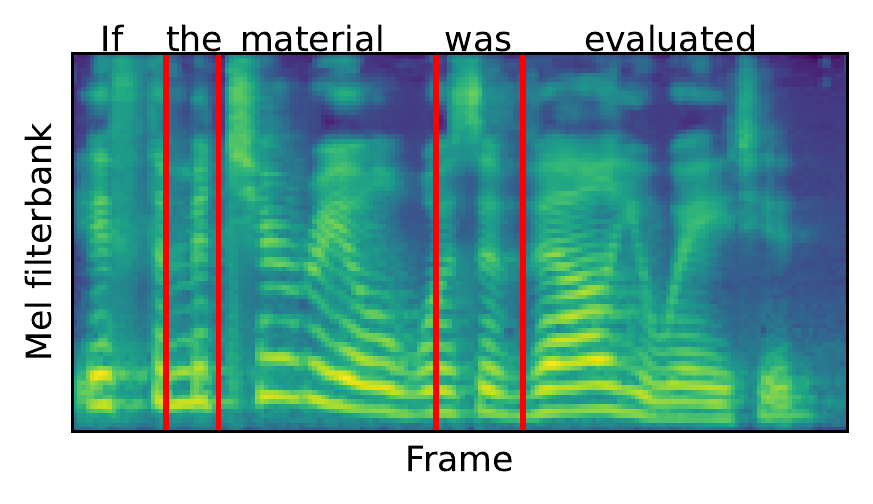}}
     \hfill
     \vspace{-0.2cm}
    \caption{Alignments of \model{} during training: (a) the attention matrix and (b) word-level duration}
     \label{fig:aligment}
     \vspace{-0.5cm}
\end{figure}

For illustrative purpose, Fig.~\ref{fig:alignment:att} depicts the alignment matrix learned by \model{} for an utterance. As can be seen, our method can converge to a hard and monotonic alignment.  In addition, Fig.~\ref{fig:alignment:mel} shows the durations extracted by \model{} at word level. There is a clear distinction for each word.  It is important to emphasize that the duration of an individual phoneme can also be controlled by scaling the duration predictions at inference.

\section{Conclusions}
In this paper, we have proposed \model{}, a parallel end-to-end TTS method, which enables high-quality speech generation. Instead of relying on external aligners or teacher-student distillation techniques, \model{} can learn the alignments between text and speech from raw data, making the training pipeline simpler. We have shown that \model{} is fast at inference, while being competitive in terms of speech quality to other state-of-the-art TTS systems. For future work, we will further reduce the model size, while keeping the audio quality to ensure fast speech synthesis on small devices. Another promising direction is to apply our duration-based alignment mechanism to other domains.

\bibliographystyle{IEEEbib}
\small
\bibliography{mybib}
\end{document}